\documentclass[pra,preprint,superscriptaddress]{revtex4-1}

\usepackage{amsmath}    % need for subequations
\usepackage{graphicx}   % for figures
\usepackage{color}
\usepackage[10pt]{moresize}

\usepackage{amsfonts}
\usepackage{amssymb}
\usepackage{amscd}
\usepackage{enumerate}
\usepackage{epsfig}
\usepackage{subfigure}
\usepackage{graphicx}
\usepackage{bm}
\usepackage{color}

\begin{document}

\title{Direct Counterfactual Communication via Quantum Zeno Effect}

\author{Yuan Cao}
\thanks{These authors contributed equally to this work}
\author{Yu-Huai Li}
\thanks{These authors contributed equally to this work}
\affiliation{Shanghai Branch, National Laboratory for Physical Sciences at Microscale and Department of Modern Physics, University of Science and Technology of China, Shanghai 201315, China}
\affiliation{Synergetic Innovation Center of Quantum Information and Quantum Physics, University of Science and Technology of China, Hefei, Anhui 230026, China}
\author{Zhu Cao}
\affiliation{Center for Quantum Information, Institute for Interdisciplinary Information Sciences, Tsinghua University, Beijing 100084, China}
\affiliation{Synergetic Innovation Center of Quantum Information and Quantum Physics, University of Science and Technology of China, Hefei, Anhui 230026, China}
\author{Juan Yin}
\author{Yu-Ao Chen}
\author{Hua-Lei Yin}
\author{Teng-Yun Chen}
\affiliation{Shanghai Branch, National Laboratory for Physical Sciences at Microscale and Department of Modern Physics, University of Science and Technology of China, Shanghai 201315, China}
\affiliation{Synergetic Innovation Center of Quantum Information and Quantum Physics, University of Science and Technology of China, Hefei, Anhui 230026, China}
\author{Xiongfeng Ma}
\affiliation{Center for Quantum Information, Institute for Interdisciplinary Information Sciences, Tsinghua University, Beijing 100084, China}
\affiliation{Synergetic Innovation Center of Quantum Information and Quantum Physics, University of Science and Technology of China, Hefei, Anhui 230026, China}
\author{Cheng-Zhi Peng}
\email{pcz@ustc.edu.cn}
\author{Jian-Wei Pan}
\email{pan@ustc.edu.cn}
\affiliation{Shanghai Branch, National Laboratory for Physical Sciences at Microscale and Department of Modern Physics, University of Science and Technology of China, Shanghai 201315, China}
\affiliation{Synergetic Innovation Center of Quantum Information and Quantum Physics, University of Science and Technology of China, Hefei, Anhui 230026, China}

\begin{abstract}
Intuition from our everyday lives gives rise to the belief that information exchanged between remote parties is carried by physical particles.
Surprisingly, in a recent theoretical study [Salih H, et al. (2013) Phys. Rev. Lett. 110:170502], quantum mechanics was found to allows for communication even without the \emph{actual} transmission of physical particles.
From the viewpoint of communication, this mystery stems from a (non-intuitive) fundamental concept in quantum mechanics --- wave-particle duality.
All particles can be described fully by wave functions.
To determine whether light appears in a channel, one refers to the amplitude of its wave function.
However, in counterfactual communication, information is carried by the phase part of the wave function.
Using a single-photon source, we experimentally demonstrate the counterfactual communication and successfully transfer a monochrome bitmap from one location to another by employing a nested version of the quantum Zeno effect.
\end{abstract}

\maketitle
The concept of \emph{counterfactuality} originated from interaction-free measurements that were first presented in 1981 \cite{Dicke:Interaction:1981,Elitzur:interactionfree:1993}, in which the achievable efficiency was limited by a margin of 50\%.
Interaction-free measurements display a surprising consequence, wherein the presence of an obstructing object (acting as a measuring device placed in an interferometer) can be inferred without the object directly interacting with any (interrogating) particles.
Later, the efficiency was improved to 100\% \cite{Zeilinger:interactionfree:1995} by using the quantum Zeno effect \cite{Misra:quantumzeno:1977,peres:zeno:1980,Agarwall:opticalzeno:1994}, wherein a physical state experiences a series of weak measurements.
When the measurements are weak enough, the state is frozen in its initial state with a high probability.
This scheme was later applied to quantum interrogation \cite{Kwiat:Zeno:1999}, quantum computation \cite{Kwiat:cfcomputation:nature2006,PhysRevLett.115.080501} and quantum cryptography \cite{Noh:counterfactualQC:2009,Ren:ExpCounterfacrualQKD:LaserPhysics2011, Brida:ExpCounterfacrualQKD::LPL2012,liuyang:counterfactualQKD:PRL2012}.
It can also be used for creating entanglement between distant atoms \cite{arxiv1610.07169,Elitzur:2014fz}.
As for communication, counterfactuality refers specifically to cases in which information is exchanged without physical particles traveling between two remote parties.
Unfortunately, none of the previous schemes can be used for direct counterfactual communication because particles would appear in the channel for at least one logic state when information is transmitted.

However, a breakthrough in direct counterfactual quantum communication was made by Salih, Li, Al-Amri and Zubairy (SLAZ) \cite{Zubairy:counterfactualQC:2013} to solve this challenge, which raised a heated debate on its interpretation and on whether full counterfactuality can be maintained when a blockade is absent within a channel \cite{PhysRevLett.111.240402,PhysRevLett.112.208901,PhysRevLett.112.208902,PhysRevA.88.046102,PhysRevA.88.046103,PhysRevA.89.033825,LF50303,1751-8121-48-46-465303,PhysRevA.92.052315,PhysRevA.93.066301,PhysRevA.93.066302,arXiv1605.02181}.
Although several publications are presently available regarding the theoretical aspects of the subject, a faithful experimental demonstration is however missing.
Here, by employing the quantum Zeno effect and a single-photon source, a direct communication without carrier particle transmission --- the SLAZ scheme---has been successfully implemented.

First, we review the SLAZ scheme \cite{Zubairy:counterfactualQC:2013} briefly.
The first ingredient is a tandem interferometer that uses a large number ($M$) of beam splitters (BSs) with a very high reflectivity of $cos^{2}(\pi /2M)$.
Two single-photon detectors (SPDs) are placed in the two output ports of the last BS.
According to the quantum Zeno effect, one can predict which SPD would click upon Bob's choice of either blocking the upper-side arms or allowing the photons to pass.
In general, as $M$ approaches infinity, the resulting success probability approaches 100\%.
To realize direct counterfactual communication, however, this is insufficient because the photon may travel through the channels when Bob allows them to pass.
The second ingredient is the chained quantum Zeno effect, which is at the core of the SLAZ scheme.
In each of $M$ outer cycle's arms, an additional tandem interferometer is nested using $N$ BSs with a reflectivity of $cos^{2}(\pi /2N)$ to form the inner cycles.
Hence, there are in total $(M-1)(N-1)$ interferometers in the scheme.
By combining the two abovementioned ingredients, complete counterfactuality can be achieved in direct communication.
That is, when $M$ and $N$ approach infinity, the probability of photons showing up in the transmission channel approaches zero.
Therefore, the SLAZ scheme requires an infinite number of tandem interferometers, which is obviously impractical.
Moreover, in practice, total visibility deteriorates exponentially with the number of interferometers employed.
Here, we simplify the SLAZ scheme while preserving its counterfactual properties via a nested polarization Michelson interferometer and a heralded single-photon source.

\begin{figure}%[tbhp]
\centering
\includegraphics[width=1.0\linewidth]{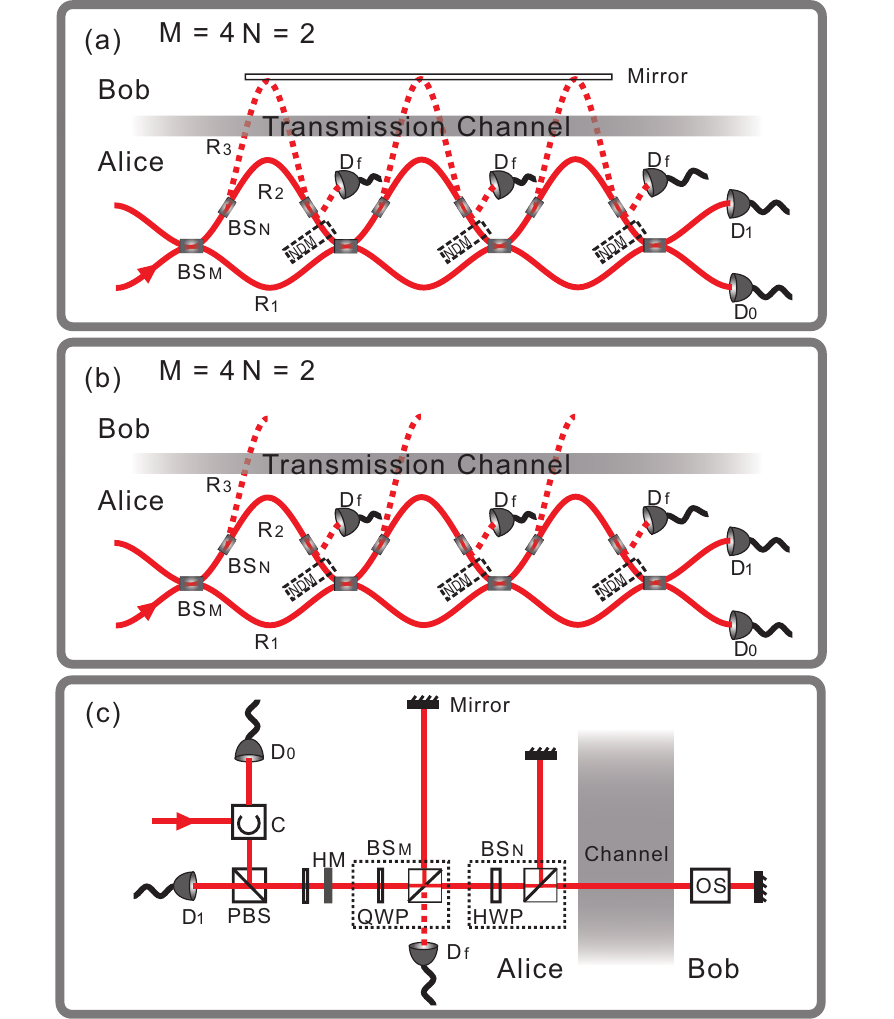}
\caption{\textbf{
Schematic diagram of direct counterfactual quantum communication. 
a}, Schematic diagram of simplified SLAZ scheme.
Biased BSs are used to obtain a certain reflection.
Bob switches between PASS and BLOCK by inserting or removing the mirror.
Two types of $BS$ are used: $BS_M$ with transmittance of ${\sin ^2}(\pi /2M)$ and $BS_N$ with transmittance of ${\sin ^2}(\pi /2N)$.
The three paths, R1, R2, and R3, are of the same length.
\textbf{b}, With the mirror in \textbf{a} removed, Route 3 is broken.
\textbf{c}, Experimental implementation of simplified SLAZ scheme by employing a nested polarization Michelson interferometer.
The biased BSs are realized by PBSs and wave plates, a simple and accurate method to control reflection.
In our experiment, $M=4$ and $N=2$.
The optical switch (OS) in \textbf{c} corresponds to the removable mirror in \textbf{a} and \textbf{b}.
QWP: quarter-wave plate; HWP: half-wave plate; C: circulator; HM: half mirror; R1, R2, and R3: Route 1, Route 2, and Route 3; NDM: Nondemolition measurement.}
\label{Fig:zeno:theoryfig}
\end{figure}

A schematic of the simplified SLAZ scheme is shown in Fig.~\ref{Fig:zeno:theoryfig}a.
A single photon is transferred by Alice to the nested interferometer and is detected subsequently by three individual single-photon detectors, $D_0$, $D_1$ and $D_f$.
Alice concludes a logic result of either 0 or 1 depending on whether detector $D_0$ or $D_1$ clicks, respectively.
Otherwise, if detector $D_f$ clicks, Alice obtains an inconclusive result, which is discarded in the data post-processing phase.
As a result of BS transformations, there are three potential routes, namely, Route 1, Route 2, and Route 3, corresponding, respectively, to the lower-side arms of the outer cycles, the lower-side arms of the inner cycles and the upper-side arms of the inner cycles, as shown in Fig.~\ref{Fig:zeno:theoryfig}a.

In the case of the logic 0, Bob emplaced mirrors in the corresponding required positions to ensure that the inner interferometer works, as shown in Fig.~\ref{Fig:zeno:theoryfig}a.
Within the domain of an infinite $M$ and ideal interference, a single photon will go to $D_0$ with a probability equal to one (i.e. without transmitting through the channel).
A finite $M$, however, may cause an erroneous event, where $D_1$ clicks for the logic 0 state.
Moreover, a finite $M$ allows a photon to pass through the channel with a nonzero probability, in which case, owing to the interference of Route 2 and 3, the photon can only be detected by $D_f$.
In the case of successful information transfer from Bob to Alice, no photon will pass through the transmission.
That is, when single photons are used, the counterfactual property is preserved in the case of logic 0 for a finite $M$ and $N$.

In the case of logic 1, Bob removes the mirrors, resulting in breaking of the inner interferometer cycle, as shown in Fig.~\ref{Fig:zeno:theoryfig}b.
In this case, counterfactuality is guaranteed by the structure of the outer interferometer and is thereby not dependent on the value of $M$ or $N$.
The transmission channel is broken, and hence, any detection on Alice's side is not caused by photons transmitted through the channel (even if $N$ is small).
That is, the counterfactual property is preserved for the case of logic 1 in all practical scenarios.
If $N$ approaches infinity and the outer interferometer is ideal, the probability of a single photon going to $D_1$ approaches 100\%.
An imperfect outer interferometer, however, may cause an erroneous event where $D_0$ clicks for the logic 1 case.

Since the presence of any photon in the channel would lead to detection events at $D_{f}$ (logic 0, PASS) or at Bob's optical switch (logic 1, BLOCK), no photons pass through the transmission channel (Route 3) when Alice can learn the logic state (pass or block) of Bob's setting.
Considering that the errors of logic 0 are related only to the value of $M$, we choose $M = 4$ for the outer loop and $N = 2$ for the inner loop (more details on this choice in Materials and Methods).
Note that the single-photon source used in the modified SLAZ scheme cannot be replaced by a strong coherent state (classical) light.
Otherwise, because the number of interferometers is finite in the modified SLAZ scheme, a coherent light will pass through the channel with nonzero amplitude even when one of Alice's detectors clicks (which violates the counterfactual property).
More precisely, given that the reflectivity of the BS is less than 100\%, a portion of the coherent light will be transmitted through the channel with nonzero amplitude, while the rest of it will reach Alice's detectors with nonzero amplitude.
Thus, owing to multiple photons, post-selection via one of Alice's detectors clicking cannot ultimately prevent light transmission through the channel with nonzero amplitude.

\begin{figure*}[!t]\center
\resizebox{14cm}{!}{\includegraphics{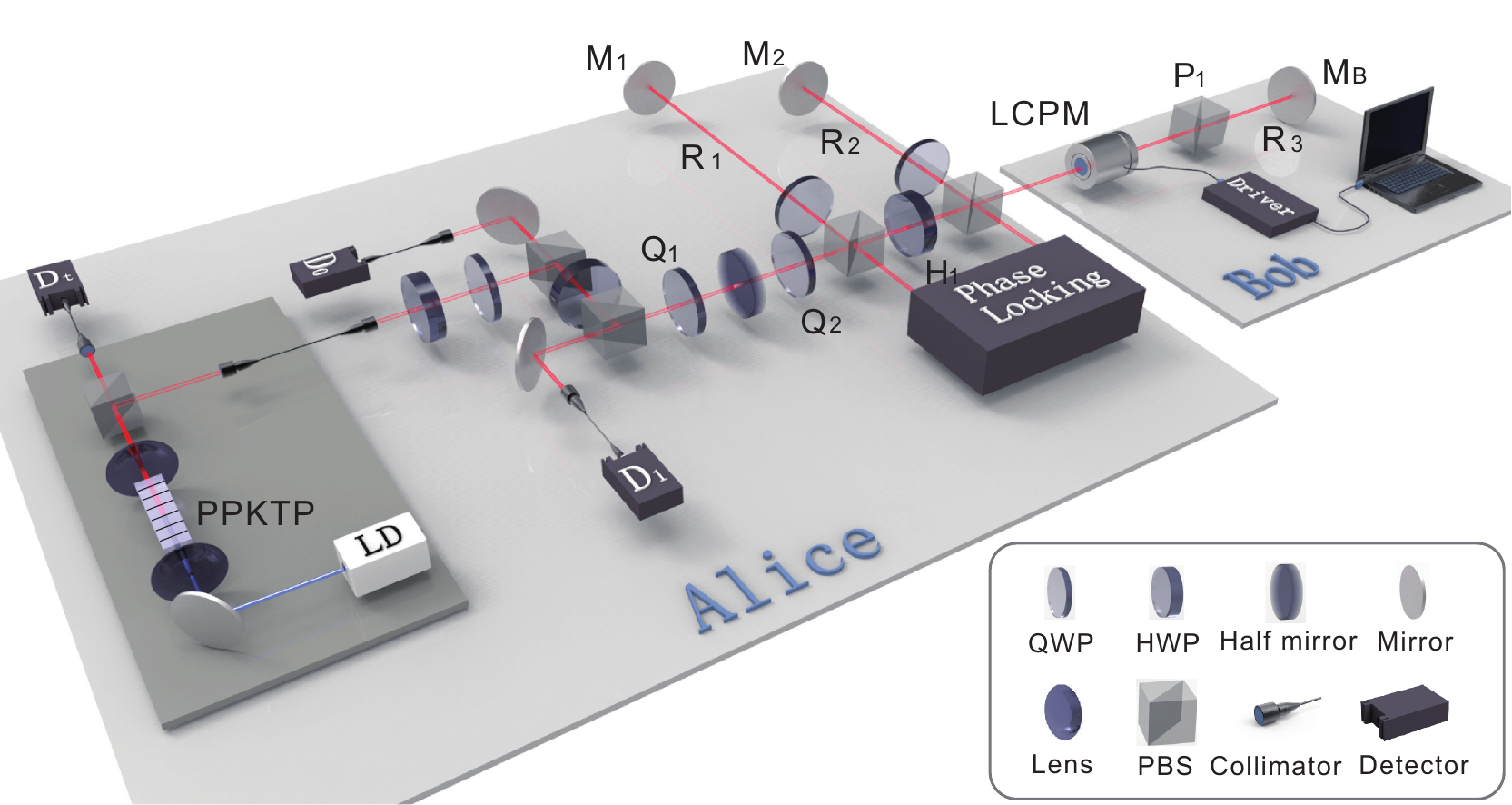}}
\caption{(color online). \textbf{
Sketch of experimental setup}.
An SPDC source is employed as a heralded single-photon source.
The two optical paths $R_1$ and $R_2$ correspond to the outer cycle of the nested Zeno effect in the Michelson interferometer setup, while the paths $R_2$ and $R_3$ (the transmission channel) correspond to the inner cycle interferometer.
Alice and Bob’s boxes are separated by 50 cm.
$D_0$, $D_1$, and $D_t$: single-photon detectors; QWP: quarter-wave plate; HWP: half-wave plate; PBS: polarizing beam splitter, reflecting vertically polarized photons; LCPM: liquid crystal phase modulator.}
\label{Fig:cfcomm:setup}
\end{figure*}

In our experimental implementation, a Michelson-type configuration, shown in Fig.~\ref{Fig:zeno:theoryfig}c is employed, which is equivalent to Fig.~\ref{Fig:zeno:theoryfig}a and b.
The experimental setup of counterfactual communication is shown in Fig.~\ref{Fig:cfcomm:setup}.
A heralded single-photon source based on a spontaneous parametric down-conversion (SPDC) process is used on Alice's side.
Additional details on the source are given in Materials and Methods.
The generated photon pairs are coupled into two single-mode fibers.
In the heralding arm, the photon enters the detector $D_t$ directly, whose timing is recorded by a high-speed and high-accuracy time-to-digital converter (TDC).
In the signal arm, the photon transitioned into a concatenation of two polarized Michelson interferometers via a collimator for direct counterfactual communication.

The proportion of multi-photon components in the signal arm is about 1.8\% under a coincidence window of 1 ns.
Although the multi-photon components do not lead to an increase in the error rate, they do attenuate tangibly the counterfactual property when a finite number of imperfect interferometers are used, which is similar to the case of using coherence state lights for counterfactual communication (for details, see Materials and Methods).
With a 10 dB collection loss in the heralded single-photon source, the multi-photon probability is reduced to an extremely low level, which could be ignored in practice.

The signal photon needs to be controlled precisely to pass the nested Michelson interferometer thrice to ensure $M = 4$ for the outer loop, which can essentially be realized as follows.
Step 1, a mirror is placed at the entrance of the outer interferometer after photon incidence.
Step 2, a photon oscillates through the outer interferometer for three times.
Step 3, the mirror is then removed so that the photon can emanate from the outer interferometer for subsequent detection by $D_0$ or $D_1$.
Such a scheme \cite{Zubairy:counterfactualQC:2013} requires high-frequency emplacement and removal of the mirror so that it can match the frequency of the photon pulse.
This speed needs to be on the order of nanoseconds in our case, which is technically challenging.
As such, we employ a half-mirror strategically in place of a high-speed removable mirror, and details on the  control of $M$ can be found in Materials and Methods.

To realize active choice between the two states (Pass [logic 0] and Block [logic 1]), a liquid-crystal phase modulator (LCPM) and a polarizing beam splitter (PBS) are used on Bob's side.
If Bob were to choose logic 1, the LCPM would apply a $\pi$-phase delay on the arriving photon, which converts the polarization of photon from horizontal ($H$) to vertical ($V$).
The photon would then be reflected by $P_1$ and discarded, so that the transmission channel is blocked; otherwise, the LCPM does not affect the arriving photon.
On Alice's side, a bit 0 on the coincident detection of $D_0$ and $D_t$ and a bit 1 on the coincident detection of $D_1$ and $D_t$ are recorded.

Another challenge in realizing the nested interferometer is maintaining a high visibility level for counterfactual communication, which required stability in the sub-wavelength order.
We employ an active phase stabilization technique in the experiment to suppress mechanical vibration and temperature drift.
We replace detector $D_f$ with a phase stabilization system to run direct counterfactual communication.
Interference visibility can be maintained at 98\% for several hours, and more details on phase stabilization can be found in Materials and Methods.

\begin{figure*}[!t]\center
\resizebox{14cm}{!}{\includegraphics{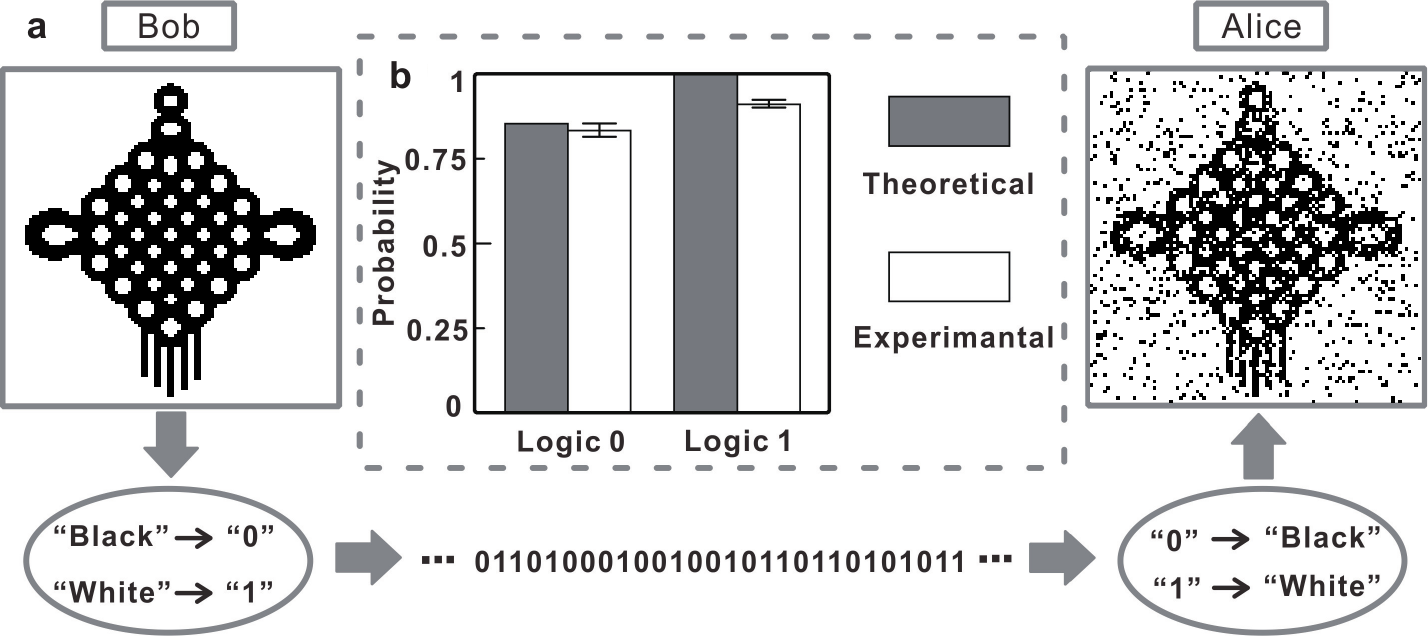}}
\caption{(color online). \textbf{
Experiment of direct counterfactual communication: transmitting a monochrome bitmap of Chinese knot.
a}, Comparison of original and transferred images.
The black pixel is defined as logic 0, while the white pixel is defined as logic 1.
\textbf{b}, Probabilities of transmitting logic $0$ and logic $1$.
Experimental results are compared with the theoretical limits.}
\label{Fig:Node:result}
\end{figure*}

Direct counterfactual communication was demonstrated by transmitting a $100\times100$ pixel monochrome bitmap (Chinese knot), as shown in Fig.~\ref{Fig:Node:result}.
Bit by bit, Bob controlled his LCPM according to 10-Kbits bitmap information processed over 5 hours with a total channel loss of 52 dB.
Because of this channel loss, unfortunately, many bits were ultimately not detected successfully.
As a result, Alice was needed to send single photons repeatedly until a successful detection event (either $D_0$ and $D_t$ clicking, or $D_1$ and $D_t$ clicking) was obtained.
Subsequently, she then regularly transmitted feedback to Bob informing him to continue with the next bit until all 10 Kbits of information were transmitted.

In the ideal case of the SLAZ scheme with $M = 4$ and $N = 2$, the probabilities of Alice identifying correctly Bob's logic $0$ and logic $1$ are 85.4\% and 100\%, respectively.
In our experiment, owing to imperfect interference of the interferometers, these probabilities are reduced to \textbf{$83.4\%\pm2.2\%$} and \textbf{$91.2\%\pm1.1\%$}, respectively.
The error rate was found to be considerably higher for Logic 1.
For cases in which Bob chooses Pass (i.e. Logic 0), errors are introduced mainly by the interference imperfections of the last inner loop.
However, when Bob chooses Block (i.e. Logic 1), the three outer loops are chained, resulting in a rapid decline in the overall visibility owing to the accumulation of errors manifesting from all three interferometers.
As shown in Fig.~\ref{Fig:Node:result}, the Chinese knot bitmap is successfully transmitted from Bob to Alice with high visibility.
In principle, one can perform error correction efforts in the data post-processing stage so that information can be transmitted reliably in a deterministic way.
Moreover, the probability of Alice identifying Bob's logic state rightfully could be improved further by increasing the number of interferometer cycles and enhancing their quality.

There are several interesting related endeavors planned for future research in this topical domain.
First, image transmission can be extended from monochrome to grayscale, the key point of which is to insert such a switch in a state between Block and Pass.
This can be realized potentially via the emplacement of a partial-pass switch.
Whether such a process of direct communication maintains counterfactuality (when a partial-pass switch is used), however, should be considered carefully.
Second, the information transmitted in our current realization of counterfactual communication is considered `classical'.
Therefore, an interesting question to ask in the future is `can counterfactual communication also transmit quantum information?'
Recently, an affirmatively-oriented answer to this question has been conjectured \cite{PhysRevA.92.052315}.
Third, counterfactuality needs to be verified further experimentally.
One possible method to verify the phenomenon indirectly is by introducing nondemolition measurement at each output port of the inner-tandem interferometer, as shown in Fig.~\ref{Fig:zeno:theoryfig}a.
If the measurement results are nil, there are no photons at the output.
One could then conclude that there must be no photons in the inner cycles and that the notion of counterfactuality is upheld.
In practice, one can also achieve this nondemolition measurement by measuring a small leaked signal (i.e., via a largely biased beam splitter at each output port of the inner-tandem interferometer) \cite{Brassard2000PNS}.
Another potential verification method would be to perform nondemolition measurements on the inner cycle paths by using nonlinear crystals \cite{LF50303}.

The mysterious phenomenon of counterfactual communication can also be potentially understood from the imaging point of view.
Traditionally, a typical photography tool, such as a camera, records different light intensities that contain an object's spatial information.
In the 1940s, a new imaging technique --- holography --- was developed to record not only light intensity but also the phase of light \cite{Gabor:holography:nature1948}.
One may then pose the question `can the phase of light itself be used for imaging?'
The answer is \emph{yes}.
Through the demonstration of counterfactual communication, we have shown that phase can be used as an information carrier, while intensity information itself remains essentially irrelevant.
For example, assume Bob is equipped with an array of optical switches and Alice replaces her single-photon detectors with an ultra-sensitive camera.
The pattern of Bob's optical switch array can theoretically be recorded on Alice's camera without photons being transmitted through the channel.
We call this imaging process ``counterfactual imaging'' or ``phase imaging''.
Such a technique might be useful in a variety of practical applications, such as imaging ancient arts where shining a light directly is not permitted.

\begin{acknowledgments}
We acknowledge insightful discussions with H.~Salih, Q.~Zhang, Z.~Zhang, B.~Zhao and X. Yuan. This work has been supported by the National Basic Research Program of China Grants No.~2011CB921300, No.~2013CB336800, No.~2011CBA00300 and No.~2011CBA00301, the National Natural Science Foundation of China Grants No.~61073174, No.~61033001, No.~11304302, and No.~61061130540, and the Strategic Priority Research Program on Space Science, the Chinese Academy of Sciences.
\end{acknowledgments}

\section*{Methods}
\subsection{Evidence of counterfactuality}
In a practical situation, for finite $M$ and $N$, counterfactuality is guaranteed by two factors: high visibility of interference and low probability of $D_f$ and $D_0$ ($D_1$) clicking synchronously.
In the case of logic 0, photons are allowed to pass through the channel, but they can be detected only by $D_f$ owing to interference of the inner loop.
In the case of logic 1, any detection of Alice's side is not caused by photons transmitted through the channel because the transmission channel is altogether broken.
First, in our experiment, the visibility of interference was maintained at 98\% for several hours by employing active phase stabilization.
Second, an SPD is placed at one of the $BS_M$ ports, as shown in Fig.~\ref{Fig:zeno:theoryfig}b.
The conditional detection of $D_f$ on $D_0$ ($D_1$) was examined for a heralded single-photon source, which produces a result of 1.37\% (1.43\%).
The conditional detection rates were normalized by channel losses and detection efficiencies.
These two detection rates indicate how well the counterfactual property is preserved. 
To demonstrate the counterfactual property experimentally, we compare two numbers, namely, the maximal data transmission rate allowed by the leaked photons and the real transmission rate achieved by our experiment.
Among all post-selected detection events, around 98.6\% of photons did not gone through the channel, as required for the counterfactual communication.
According to the channel capacity theory, the maximal data rate that can be transmitted is 1 bit per photon detection for the specific setting in the experiment.
Thus, the maximal data transmission rate due to the leaked 1.4\% photons is 0.014 per detection.
In our experiment, on the other hand, we achieved 0.83 bit per detection (calculated by the average error rate of the data transmission, which is 12.7\%), which is significantly higher than 0.014.
Therefore, we conclude that the counterfactual property was demonstrated in our experiment.

\subsection{Active phase stabilization}
An additional phase-stabilization laser with the same wavelength as that of the single-photon source was coupled with the inner and the outer interferometers.
Mirrors $M_1$ and $M_B$ were placed on two piezoceramic translation stages that could adjust precisely the interferometers on the order of several nanometers, in accordance with existent feedback signals.
Without phase stabilization, the relative phase of the two interfering routes fluctuates dramatically, so the maximum and the minimum were flipped within a few minutes, resulting in negative visibility.
The visibility could be maintained at 98\% for several hours with phase stabilization, which shows the effectiveness of such a stabilization system.

\subsection{Heralded single-photon source}
A continual-wave ultraviolet laser (405 nm, 16 mW) was used to pump type-II periodically-poled potassium titanyl phosphate (PPKTP), creating a pair of photons in the state~$\left| {HV} \right\rangle$.
This type of SPDC source yields about $2\times 10^7$ pair/s of photons at 810 nm.
The emitted photon pairs are split into two spatial modes by a PBS, which reflects only vertically polarized photons, namely, heralding arm and signal arm respectively.
For the heralding arm, the efficiencies of fibre-coupling and detection are about 30\% and 60\%, respectively, with an overall heralding efficiency around 18\%.
As a result, the effective brightness of the heralding single-photon source is about $3.6\times 10^6$/s.

\subsection{Choosing and controlling proper $M$ and $N$}
Using a large $M$ lowers the error rate of logic 0, which is given by $1-{\cos ^2}(\pi /2M)$.
A small $N$, however, will not increase the error rate with the presence of good interference visibility.
On the other hand, photons need to travel the transmission channel $2(M-1)(N-1)$ times before detection in the counterfactual communication.
Hence, channel-loss will increase with the value of $M$ and $N$.
In addition, a chained interferometer is more difficult to stabilize with larger $M$ and $N$ values.
Therefore, in the experiment, we pick reasonable values of $M = 4$ and $N = 2$ for demonstrating direct counterfactual communication.

The desired transmission path of a photon (to be post-selected) is described as follows.
Firstly, the photon is transmitted through a half-mirror to enter a nested interferometer. Then, the photon bounces back and forth within the interferometers $M-1$ times; that is, it is reflected from the half-mirror $M-2$ times.
Finally, the photon is transmitted through the half-mirror again and reaches the detection.
Suppose the reflectivity of the half-mirror is $R$, and its transmittance is $(1-R)$, ignoring absorption.
Therefore, the probability of a photon traveling through the desired path is $(1-R)^2 R^{M-2}$, which is maximized at $R = (M-2)/M$.
The optimal reflectivity is therefore $R=50\%$ for $M = 4$.
Meanwhile, it is not difficult to see that the major penalty for such a replacement is the extra loss introduced by the half-mirror, which is given by $1-(1-R)^2 R^{M-2}$.
Hence, this loss is equivalent to 15/16 in such a configuration, corresponding to 12 dB.

As discussed previously, with a theoretical removable mirror replaced by a half-mirror, an ongoing challenge in this experiment is ensuring that the exit photons travel through the interferometer exactly three times ($M=4$).
We distinguish between desired and undesired photons by using spatial and timing modes.
Accordingly, the half-mirror is tilted at a very small angle of around $500 \mu rad$ to separate the photons experiencing different interferometer cycles based on their angles of emergence.
Only the photons that travel through the desired path are in the correct spatial mode to be coupled successfully with the single-mode fibers in front of $D_0$ and $D_1$.
Meanwhile, the time delay between the photon triggers from $D_t$ and the detection clicks from $D_0$ and $D_1$ to select the desired events.

\subsection{Realization of BSs with specified reflectivity}
Building biased BSs with specified reflectivity is a challenging work.
Alternatively, according to the SLAZ scheme \cite{Zubairy:counterfactualQC:2013}, we employ a wave plate and a PBS to realize the function of a biased BS, as shown in Fig.~\ref{Fig:cfcomm:setup}.
We aligned the optical axes of two quarter-wave plates, $Q_1$ and $Q_2$, to $\pi/16$ for $M = 4$, and that of a half-wave plate $H_1$ to $\pi/8$ for $N = 2$, as shown in Fig.~\ref{Fig:cfcomm:setup}.
The precision of wave plates alignment is below $0.5^{\circ}$.

\bibliography{pnas-sample}

\end{document}